\documentclass[aps,pre,amsmath,amssymb,twocolumn,floatfix]{revtex4}

\usepackage{graphicx}
\usepackage{xcolor}

\newcommand{\iu}{\mathrm{i}} % imaginary unit
\DeclareMathOperator{\Op}{Op}
\DeclareMathOperator{\sech}{sech}
\begin{document}

%%%%%%%%%%%%%%%%%%%%%%%%%%%%%%%%%
\title{Transparent boundary conditions for the nonlocal nonlinear Schr\"{o}dinger equation:\\ A model for reflectionless propagation of PT-symmetric solitons}
\author{M.E. Akramov$^{1}$, J.R. Yusupov$^{2}$, M. Ehrhardt$^{3}$, H. Susanto$^{4}$ and D.U. Matrasulov$^{5}$}
\affiliation{$^1$National University of Uzbekistan, 4 Universitet Str., 100174, Tashkent, Uzbekistan\\
$^2$Yeoju Technical Institute in Tashkent, 156 Usman Nasyr Str., 100121, Tashkent, Uzbekistan\\
$^3$Bergische Universit\"at Wuppertal, Gau{\ss}strasse 20, D-42119 Wuppertal, Germany\\
$^4$Khalifa University, 127788, Abu Dhabi, United Arab Emirates\\
$^5$Turin Polytechnic University in Tashkent, 17 Niyazov Str., 100095, Tashkent, Uzbekistan}

%%%%%%%%%%%%%%%%%%%%%%%%%%%
\begin{abstract}
We consider the problem of reflectionless propagation of PT-symmetric solitons described by the nonlocal nonlinear Schr\"odinger equation on a line in the framework of the concept of transparent boundary conditions for evolution equations.
Transparent boundary conditions for the nonlocal nonlinear Schr\"odinger equation are derived.
The absence of backscattering at the artificial boundaries is confirmed by the numerical implementation of the transparent boundary conditions.
\end{abstract}

\maketitle

%%%%%%%%%%%%%%%%%%%%%%%%%%%%%%%%%
\section{Introduction}
Modeling of wave dynamics in various media is of practical importance in the fields of optics, optoelectronics, fluid dynamics, acoustics and communication technology. 
An important problem to be solved by such models is tunable wave propagation, which means achieving ballistic and diffusive states or absence of backscattering and reflection in a certain subdomains. 
In quantum mechanics, such a goal can be achieved by constructing a suitable scattering matrix that ensures the absence of reflection. 

However, in the case of nonlinear wave propagation, one cannot use a scattering matrix and must develop efficient mathematical tools to describe reflectionless propagation.
One of these tools can be based on the use of the concept of so-called \textit{``transparent boundary conditions (TBCs)''} (other names are ``artificial boundary conditions'' and ``absorbing boundary conditions''). 
The concept was previously applied to the linear \cite{Ehrhardt1999,Ehrhardt2001,Arnold2003,Ehrhardt2008} and nonlinear \cite{Antoine,Han0,Matthias2008,Antoine2008,Zhang} Schr\"odinger equations. 
Recently, the application of the TBC concept to Dirac-\cite{Hammer2014} and Klein-Gordon equations \cite{Gander,Antoine1}, including the nonlinear Klein-Gordon equation \cite{Zheng07,Han,Li,Antoine1,TBCSGE} was shown.
Extensions of TBCs for evolution equations on graphs can be found in references \cite{Jambul,Jambul02,Jambul1,Jambul2,Jambul4}. 

The basic idea of the TBC concept can be formulated as follows: 
For a given partial differential equation formulated as an initial value problem in a finite domain, it is required that the solution in one domain should match that in the entire space restricted to the finite domain. 
This can be achieved by artificial boundary conditions, which have a rather complicated form (e.g., for Schr\"odinger or Dirac equations given in terms of fractional derivatives are \cite{Ehrhardt1999,Hammer2014}).

In this paper we extend the TBC concept for the \textit{nonlocal nonlinear Schr\"odinger equation (NNLS)} describing the dynamics of PT-symmetric solitons. 
The NNLS equation was first introduced by Ablowitz and Musslimani \cite{AM2013}, who showed the integrability of the problem and obtained their soliton solutions. 
Later it was used in various contexts in Refs.
\cite{AM2013,Stalin,AM2014,AM2016,AM20161,Sinha,Yang,Zhenya,AM2018,AM2018_1,AM2019,Hadi2019,Kanna2020,Panos2020,Mashrab2022}. 
The NNLS equation describes the dynamics of solitons in media with self-induced PT-symmetric nonlinearity (such nonlinearity may be present, for example, in an optical waveguide with self-induced gain loss). 
Our proposed model accounts for reflectionless propagation of solitons in such media. 
The transparent boundary conditions derived here provide mathematical constraints that ensure that there is no backscattering. 
In practical applications of boundary conditions, e.g., for branched waveguides, they can provide physically acceptable constraints on the equivalence of the usual weight continuity and Kirchhoff's rules with the transparent boundary conditions at the branching point (see, e.g., Refs.~\cite{Jambul,Jambul02,Jambul1} for details). 

This paper is organized as follows. 
In the next section, we briefly recall soliton solutions and conserving quantities for the NNLS equation on a line. 
In Section~III, we derive the transparent boundary conditions for the NNLS equation. 
In Section~IV, we demonstrate our numerical implementation of such boundary conditions and show the results of numerical experiments in Section~V. 
Finally, Section~VI contains the concluding remarks.

%%%%%%%%%%%%%%%%%%%%%%%%%%%%%%%%%%%%%%
\section{Soliton solutions of the nonlocal nonlinear Schr\"{o}dinger equation}

Here we briefly recall basic results on the NNLS equation on a line, following Ref.~\cite{AM2013}.
The nonlocal nonlinear Schr\"odinger equation is given as \cite{AM2013}
\begin{equation}\label{nnlse}
    \iu\partial_t q(x,t)+\partial^2_x q(x,t)+2 q(x,t) q^*(-x,t) q(x,t)=0, 
\end{equation}
where $q^*$ denotes the complex conjugate of $q$ and the potential, which can be defined as $V(x,t)=2\,q(x,t)\,q^*(-x,t)$, has the PT symmetric property, i.e.\ $V(x,t)=V^*(-x,t)$. 
We note that the nonlocality of Eq.~\eqref{nnlse} results from the fact that the evolution of the field $q(x,t)$ at coordinate $x$ always requires information from the opposite point $-x$. 
For the above NNLS equation, there are many different types of soliton solutions, namely breathing, periodic, rational, and others. 
A single soliton solution can be found by the inverse scattering method as given in Ref.~\cite{AM2013}:
\begin{equation}\label{sol01} 
    q(x,t)=-\frac{2(\eta_1+\bar{\eta}_1)\,e^{\iu\bar{\theta}_1}\,e^{4\iu\bar{\eta}^2_1 t}\,e^{-2\bar{\eta}_1x}}{1+e^{\iu(\theta_1+\bar{\theta}_1)} \,e^{-4\iu(\eta^2_1-\bar{\eta}^2_1)t} \,e^{-2(\eta_1+\bar{\eta}_1)x}},
\end{equation}
with $\eta_1$, $\bar{\eta}_1$, $\theta_1$, and $\bar{\theta}_1$ being real constants. 
The traveling soliton solution of Eq.~\eqref{nnlse} can be written as \cite{Stalin}
\begin{gather} \label{travelling}
    q(x,t)=\frac{\alpha_1 \,e^{-\Delta/2}\,
    e^{(\bar{\xi}_{1R}-\xi_{1R})+i(\bar{\xi}_{1I}-\xi_{1I})}   }{2[\cosh(\chi_1)
    \cos(\chi_2) + \iu \sinh(\chi_1)\sin(\chi_2)]  },
\end{gather}
where $\xi_{1R}=-k_{1I} (x+2k_{1R}t)$, $\xi_{1I} = k_{1R} x - (k_{1I}^2 - k_{1R}^2) t$, 
$\bar{\xi}_{1R}=-\bar{k}_{1I} (x+2\bar{k}_{1R}t)$, $\bar{\xi}_{1I} = \bar{k}_{1R} x - (\bar{k}_{1R}^2 - \bar{k}_{1I}^2) t$, $\Delta_R = \log\left(\frac{|\alpha_1|^2 |\beta_1|^2}{|k_1+\bar{k}_1|^2}\right)$, 
$\Delta_I = -\frac{i}{2} \log\left(  \frac{\alpha_1 \beta_1 (k_1^*+\bar{k}_1^*)^2}{\alpha^*_1 \beta^*_1 (k_1+\bar{k}_1)^2} \right)$, $e^{\Delta} = -\frac{\alpha_1 \beta_1}{(k_1+\bar{k}_1)^2}$, $\chi_1=(\xi_{1R}+\bar{\xi}_{1R}+\Delta_R)/2$ and $\chi_2=(\xi_{1I}+\bar{\xi}_{1I}+\Delta_I)/2$.

The integrability of the problem was proved in \cite{AM2013}, which means that the NNLS equation has many conservation laws.
In particular, two important conservation quantities, the norm and the energy, were derived in \cite{AM2013} and can be written as
\begin{align}\label{energy01}
    N(t)=\underset{-\infty}{\overset{+\infty}{\int}}q(x,t)q^*(&-x,t)\,dx,\nonumber\\
    E(t)=\underset{-\infty}{\overset{+\infty}{\int}}\Big[\frac{\partial}{\partial x}q(x,t)&\cdot\frac{\partial}{\partial x}q^*(-x,t)\nonumber\\
&+q^2(x,t)\cdot q^{*2}(-x,t)\Big]\,dx.
\end{align}

The above soliton solutions of Eq.~\eqref{nnlse} are obtained assuming asymptotic boundary conditions at infinity, i.e.\ $q(x,t) \to 0$ at $x \to \pm\infty$. 
In the next section, we impose additional (artificial) boundary conditions for a given finite interval $[-L, L]$ that allow an almost reflection-free transmission of a soliton through the points $\pm L$.

%%%%%%%%%%%%%%%%%%%%%%%%%%%%%%%%%%%%%%%%%%%%%%
\section{Transparent boundary conditions for the nonlocal nonlinear Schr\"{o}dinger equation}

Here we consider the problem of transparent boundary conditions for the NNLS equation \eqref{nnlse}. 
To derive TBCs for the nonlocal nonlinear Schr\"{o}dinger equation, we use the so-called \textit{potential approach}, which was proposed earlier in \cite{Antoine} and used to derive TBCs for nonlinear Schr\"odinger equations \cite{Jambul3} and sine-Gordon equations \cite{TBCSGE}. 
Within the framework of this approach, the NNLS equation can be formally reduced to the linear Schr\"{o}dinger equation
\begin{equation}\label{nlse}
   \iu\partial_t q(x,t)+\partial^2_x q(x,t)+ V(x,t) q(x,t)=0, 
\end{equation}
with the potential is $V(x,t)=2 q(x,t) q^*(-x,t)$. 
In the remainder of this section, we invoke the same procedure and derive TBCs at the end. 

To do so, we introduce a new unknown $Q(x,t)$, which is given by the relation
\begin{equation}\label{Q}
    Q(x,t)=e^{-\iu \mathcal{V}(x,t)}\, q(x,t), 
\end{equation}
where 
\begin{equation}\label{potential}
     \mathcal{V}(x,t)=\int_0^t V(x,s)\,ds.  
\end{equation}

The temporal and spatial derivatives of $q$ can be written as derivatives of $Q$ as
\begin{equation}
    \partial_t q = e^{\iu\mathcal{V}} (\partial_t + \iu V )Q,    
\end{equation}
and
\begin{equation}
   \partial^2_x q = \iu e^{\iu\mathcal{V}} (\partial^2_x Q + 2\iu\partial_x \mathcal{V} \partial_x Q+ \iu Q \partial^2_x \mathcal{V} -(\partial_x \mathcal{V})^2 Q).    
\end{equation}

As a result, we obtain the Schr\"{o}dinger equation in terms of $Q(x,t)$ as
\begin{equation}\label{L_operator1}
   L(x,t,\partial_x, \partial_t)Q=\iu\partial_t Q + \partial_x^2 Q + A \partial_x Q + B Q = 0, 
\end{equation}
where $A=2\iu\partial_x \mathcal{V}$ and $B=(\iu\partial^2_x \mathcal{V} - (\partial_x \mathcal{V})^2)$.
Linearizing Eq.~\eqref{L_operator1} using the pseudo-differential operator calculus, we obtain
\begin{multline}\label{L_operator2}
L=(\partial_x+\iu\Lambda^-)(\partial_x+\iu\Lambda^+) \\
=\partial_x^2+\iu(\Lambda^+ + \Lambda_-)\partial_x + \iu \Op(\partial_x \lambda^+) - \Lambda^+ \Lambda^-, 
\end{multline}
where $\lambda^+$ denotes the principal symbol of the operator $\Lambda^+$ and $\Op(p)$ denotes the associated operator of a symbol $p$. 
From the Eqs.~\eqref{L_operator1} and \eqref{L_operator2} we obtain the system of operators
\begin{gather}
    i(\Lambda^+ + \Lambda^-) = A \nonumber\\
    i \Op(\partial_x \lambda^+) - \Lambda^+ \Lambda^- = i\partial_t + B,
\end{gather}
which yields the symbolic system of equations
\begin{gather}
\iu(\lambda^+ + \lambda^-) = a \nonumber\\
\iu\partial_x \lambda^+ -\sum_{\alpha=0}^{+\infty} \frac{(-1)^{\alpha}}{\alpha!} \partial_{\tau}^{\alpha} \lambda^- \partial_t^{\alpha} \lambda^+ = -\tau + b,\label{eq01} 
\end{gather}
where $\Op(a)=A$ and $\Op(b)=B$ can be set as $a=A$ and $b=B$, since these two functions correspond to zero-order operators. 
An asymptotic evolution in the inhomogeneous symbols is defined as
\begin{equation}\label{eq02}
     \lambda^{\pm} \sim \sum_{j=0}^{+\infty} \lambda_{1/2-j/2}^{\pm}. 
\end{equation}
Substituting the expansion \eqref{eq02} into Eq.~\eqref{eq01}, one can identify the terms of order $1/2$ in the first relation of the system \eqref{eq01}:
\begin{equation}
     \lambda_{1/2}^- = -\lambda_{1/2}^+, \quad \lambda_{1/2}^+ = \pm \sqrt{-\tau}.
\end{equation}
The Dirichlet-to-Neumann operator corresponds to the choice
$\lambda_{1/2}^+ = \pm \sqrt{-\tau}$. 
For the zeroth-order terms we obtain
\begin{gather}
  \lambda_0^- = -\lambda_0^+ -\iu a,\nonumber\\
 \iu\partial_x \lambda_{1/2}^+ - (\lambda_0^- \lambda_{1/2}^+ + \lambda_0^+ \lambda_{1/2}^-)=0.\label{eq03} 
\end{gather}
From Eq.~\eqref{eq03} we get
\begin{gather}
   \lambda_0^+ = -\iu\frac{a}{2} = \frac{1}{2} \partial_x \mathcal{V}, \nonumber\\
\lambda_0^- = -\lambda_0^+ -\iu a = \frac{1}{2} \partial_x \mathcal{V}.
\end{gather}
For the terms of order $-1/2$ we obtain
\begin{gather}\label{eq04} 
    \iu(\lambda_{-1/2}^+ + \lambda_{-1/2}^-) =  0,\nonumber\\
    \iu\partial_x \lambda_0^+ - (\lambda_{-1/2}^- \lambda_{1/2}^+ + \lambda_0^+ \lambda_0^- + \lambda_{-1/2}^+ \lambda_{1/2}^-) = b,
\end{gather}
since $\partial_t^{\alpha} \lambda_{-1/2}^{\pm} = \partial_{\tau}^{\alpha} \lambda_0^{\pm}=0$, $\alpha\in N$.
From Eq.~\eqref{eq04} we get
\begin{equation}
   \lambda_{-1/2}^{\pm}=0.
\end{equation}
Furthermore, the terms of the next order can be obtained as
\begin{equation}
    \lambda_{-1}^- = -\lambda_{-1}^+ ,
    \quad \lambda_{-1}^+ = \iu\frac{\partial_x V}{4\tau}.
\end{equation}

Thus, the first-order approximation is
\begin{subequations}\label{first_approx}
\begin{gather}
    \partial_x q|_{x=-L} - e^{-\iu\frac{\pi}{4}} e^{\iu\mathcal{V}} \partial^{1/2}_t (e^{-\iu\mathcal{V}}q)   \big|_{x=-L}=0, \label{first_approx1}\\
    \partial_x q|_{x=L} + e^{-\iu\frac{\pi}{4}} e^{\iu\mathcal{V}} \partial^{1/2}_t (e^{-\iu\mathcal{V}}q)  \big|_{x=L}=0.\label{first_approx2} 
\end{gather}
\end{subequations}

The second-order approximation reads
\begin{subequations}\label{second_approx}
\begin{gather} 
    \partial_x q|_{x=-L} - e^{-\iu\frac{\pi}{4}} e^{\iu\mathcal{V}} \partial^{1/2}_t (e^{-\iu\mathcal{V}}q) \big|_{x=-L} \nonumber\\
    - \iu\frac{\partial_x V}{4} e^{\iu\mathcal{V}} I_t(e^{-\iu\mathcal{V}}q) \big|_{x=-L} = 0,\label{second_approx1}\\
    \partial_x q|_{x=L} + e^{-\iu\frac{\pi}{4}} e^{\iu\mathcal{V}} \partial^{1/2}_t (e^{-\iu\mathcal{V}}q) \big|_{x=L} \nonumber\\ + \iu\frac{\partial_x V}{4} e^{\iu\mathcal{V}} I_t(e^{-\iu\mathcal{V}}q) \big|_{x=L}=0,\label{second_approx2}
\end{gather}
\end{subequations}

Here the operator $\partial^{1/2}_t$ denotes the fractional time derivative operator of half order given as
\begin{equation*}
   \partial_t^{1/2} f(t)=\frac{1}{\sqrt{\pi}} \partial_t \int_0^t \frac{f(s)}{\sqrt{t-s}}\,ds,
\end{equation*}
and the operator $I_t(f)$ is given as
\begin{equation*}
   I_t f(t)= \int_0^t f(s) \,ds.
\end{equation*}

%%%%%%%%%%%%%%%%%%%%%%%%%%%%%%%%%%%%%%%%%%%%%%%%%%%%%%%%%%%%%%%%%%%%%%%%
%\section{Discretization of the equation and transparent boundary conditions}
\section{A discretization scheme for the transparent boundary conditions}

In this section, we present a numerical scheme for Eq.~\eqref{nnlse} and the numerical implementation of the transparent boundary conditions \eqref{first_approx} and \eqref{second_approx}. 
We have chosen the finite difference scheme of Duran--S\'anz-Serna \cite{Duran}, 
a second order scheme based on the implicit midpoint rule
\begin{multline}\label{eq::duran}
    \iu\frac{q_j^{n}-q_j^{n-1}}{\Delta t} + D^2_x \frac{q_j^{n}+q_j^{n-1}}{2} \\ + 2 \Bigl(\frac{q_j^{n}+q_j^{n-1}}{2}\Bigr)^2 
\frac{\overline{q_{J-j}^{n}}+\overline{q_{J-j}^{n-1}}}{2} = 0,
\end{multline}
with the standard second-order difference quotient
\begin{equation}
    D_x^2 q_j^n = \frac{1}{\Delta x^2} (q_{j-1}^n-2q_j^n+q_{j+1}^n), 
\end{equation}
where $\Delta x$ and $\Delta t$ are the spatial and temporal discretization steps, respectively. 
Here $\overline{q^n_j}$ denotes the complex conjugate of $q^n_j$, 
$J$ is the number of discretized spatial steps, and $q^n_j$ denotes the approximate value of $q$ at spatial coordinate $x_j$ and time $t_n$.

We now give an effective discretization scheme
for the TBC, which is implemented together with \eqref{eq::duran}. 
We give the scheme only for $x=L$, saying that the implementation for the left-hand side (at $x=-L$) can be done in the same way.
The approximation of the fractional differential operator is given by the numerical quadrature formula \cite{Antoine}
\begin{equation}
    \partial_t^{1/2} f(t_n) \approx \sqrt{\frac{2}{\Delta t}} \sum_{k=0}^n \beta_k f^{n-k},
\end{equation}
where $\{f_n\}_{n\in N}$ is a sequence of complex values approximating
$\{f(t_n)\}_{n\in N}$ and $(\beta_k)_{k\in N}$ denotes the sequences defined by
\begin{multline}
(\beta_0, \beta_1, \beta_2, \beta_3, \beta_4, \beta_5, \dots)=\\
\Bigl(   1,-1, \frac{1}{2},-\frac{1}{2}, \frac{1\cdot 3}{2\cdot 4}, -\frac{1\cdot 3}{2\cdot 4}, \dots \Bigr).
\end{multline}
The function $\mathcal{V}(x,t)$ given by \eqref{potential} can be discretized using the trapezoidal rule as
\begin{equation}
    \mathcal{V}_j^n = \Delta t \biggl[\,   \sum_{k=1}^{n-1} V_j^k + \frac{1}{2}(V_j^0 + V_j^n)   \biggr] \quad \text{for} \;n \ge 2,
\end{equation}
with $\mathcal{V}_j^0=0$ and $\mathcal{V}_j^1=\frac{1}{2}(V_j^0 + V_j^1)$, where $V_j^n$ is defined as 
\begin{equation}\label{potential_discret}
     V_j^n = 2 q_j^n \overline{q_{J-j}^{n}}. 
\end{equation}
Next, the term $e^{\iu\mathcal{V}}$ is written as
\begin{align}\label{EE}
    P_j^n :=& \exp(\iu\mathcal{V}_j^n) \nonumber\\
    =& \exp\left\{\iu\Delta t \left[ \,  \sum_{k=1}^{n-1} V_j^k + \frac{1}{2}(V_j^0 + V_j^n)   \right]\right\}. 
\end{align}
One can rewrite Eq.~\eqref{EE} as a recurrence formula
\begin{equation}
    P_j^n = P_j^{n-1} \exp\Bigl(\frac{\iu\Delta t}{2}(V_j^{n-1} + V_j^n)\Bigr).
\end{equation}
The TBC operator of the first order approximation \eqref{first_approx} on the left and right boundaries $j=0$ and $j =J$ can be approximated by the discrete convolutions
\begin{equation}
    \Lambda_1^n = e^{-\iu\pi/4} \sqrt{\frac{2}{\Delta t}} P_j^n \sum_{k=0}^n \beta_k \frac{q_j^{n-k}}{P_j^{n-k}} .
\end{equation}
Then the values of the wave function at the boundaries (together with other intermediate values) can be obtained by solving the system of nonlinear equations as
\begin{subequations}\label{eq::dappr1}
\begin{align}
\frac{q_1^n-q_{0}^n}{\Delta x} - e^{-\iu\pi/4} \sqrt{\frac{2}{\Delta t}} P_0^{n}
\sum_{k=0}^n 
\Bigl(\beta_k\frac{q_0^{n-k}}{P_0^{n-k}} \Bigr) &= 0, \label{eq::dappr1a}\\
\frac{q_J^n-q_{J-1}^n}{\Delta x} + e^{-\iu\pi/4} \sqrt{\frac{2}{\Delta t}} P_J^{n}
\sum_{k=0}^n 
\Bigl(\beta_k\frac{q_J^{n-k}}{P_J^{n-k}} \Bigr) &= 0,\label{eq::dappr1b}
\end{align}
\end{subequations}
and the Eq.~\eqref{eq::duran} with respect to $q_j^n$. 
Note that the values of $q_j^n$ to be found also exist in $P_j^n$ and here we have written Eqs.~\eqref{eq::dappr1} in short form.
Using the same approach, we can proceed with the discretization of the second-order approximation.
We approximate the integral term as
\begin{equation}
    I_j^n = \Delta t \biggl[\,  \sum_{k=1}^{n-1} \frac{q_j^k}{P_j^k} + \frac{1}{2} 
     \Bigl(   \frac{q_j^0}{P_j^0}+\frac{q_j^n}{P_j^n}   \Bigr)  \biggr].
\end{equation}
In the same way we construct a recurrence formula for $I_j^n$
\begin{equation}
    I_j^n = I_j^{n-1} + \frac{\Delta t}{2} \Bigl(\frac{q_j^{n-1}}{P_j^{n-1}}+\frac{q_j^n}{P_j^n}\Bigr),
\end{equation}
with $I_j^0=0$.
Then the TBC operator of the second-order approximation
\eqref{second_approx} takes the form
\begin{equation}
    \Lambda_2^n = \Lambda_1^n + \iu\frac{dV_j^n}{4} P_j^n I_j^n,
\end{equation}
where 
\begin{equation*}
   dV_j^n = \frac{2}{\Delta x} (  q_{j+1}^n \overline{q_{J-j}^{n}} - 2 q_j^n \overline{q_{J-j}^{n}} + q_j^n \overline{q_{J-j-1}^{n}} ).
\end{equation*}
Again, the values of the wave function at the boundaries can be obtained by solving the system of nonlinear equations with respect to $q_j^n$, given as
\begin{subequations}\label{eq::dappr2}
\begin{align}
\frac{q_1^n-q_{0}^n}{\Delta x} &- e^{-\iu\pi/4} \sqrt{\frac{2}{\Delta t}} P_0^{n}
\sum_{k=0}^n \Bigl(     \beta_k\frac{q_0^{n-k}}{P_0^{n-k}} \Bigr)\nonumber\\
&\qquad\qquad\qquad\qquad\qquad 
- \iu \frac{dV_0^n}{4}   P_0^{n} I_0^{n}, \\ 
\frac{q_J^n-q_{J-1}^n}{\Delta x} &+ e^{-\iu\pi/4} \sqrt{\frac{2}{\Delta t}} P_J^{n}
\sum_{k=0}^n \Bigl(     \beta_k\frac{q_J^{n-k}}{P_J^{n-k}} \Bigr)\nonumber\\
&\qquad\qquad\qquad\qquad\qquad + \iu \frac{dV_J^n}{4}  P_J^{n} I_J^{n}.
\end{align}
\end{subequations}

\begin{figure}[t!]
\includegraphics[width=90mm]{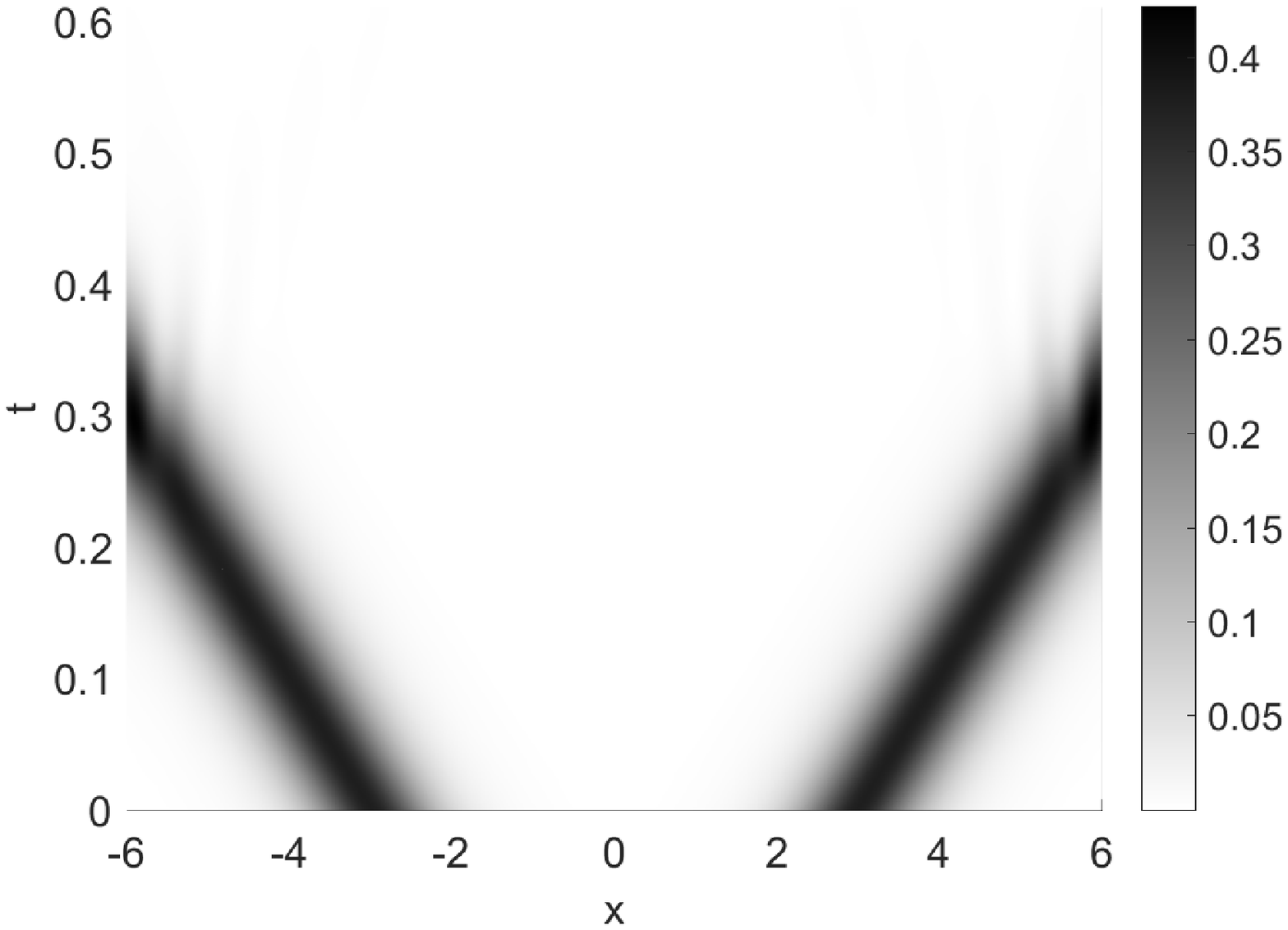}
\caption{Evolution of the soliton profile with TBCs imposed at both (left and right) end points.}
\label{fig:contour}
\end{figure}
\begin{figure}[t!]
\includegraphics[width=90mm]{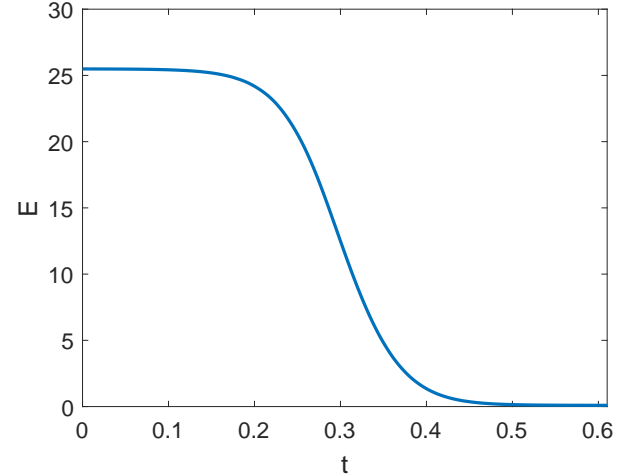}
\caption{The plot of the energy versus time in the finite domain $x\in[-L,L]$.}\label{fig:energy}
\end{figure}

%%%%%%%%%%%%%%%%%%%%%%%%%%%%%%%%%%%%%%%%%%%%%%%%%%%%%%%%%%%%%%%
\section{Numerical Experiment}
We solve the nonlocal nonlinear Schr\"odinger
equation given by Eq.~\eqref{nnlse} on the finite interval $[-L,L]$, and impose the TBC with the first-order approximation \eqref{first_approx} on the left ($x=-L$) and right ($x=L$) boundaries. 
For the initial condition, we choose the sum of two analytic solutions \eqref{travelling} that are symmetric about the origin of the spatial coordinate ($x=0$) 
\begin{equation}
     q(x,0)= \frac{\alpha_1}{2} \sqrt{\frac{3\iu}{\alpha_1 \beta_1}}\bigg(G_+(x)+G_-(x)\bigg),   
\end{equation}
where
\begin{equation*}
   G_{\pm}(x)=\frac{e^{\mp5\iu(x\pm L/2)}}{g_r^{\pm}+
    \iu g_i^{\pm}},
\end{equation*}
\begin{equation*}
    g_r^{\pm}(x)=\cosh\Bigl(\frac{-3(x\pm L/2)-\Delta_R}{2}\Bigr) \cos\Bigl(\frac{\Delta_I}{2}\Bigr),
\end{equation*}
\begin{equation*}
   g_i^{\pm}(x)=\sinh\Bigl(\frac{-3(x\pm L/2)-\Delta_R}{2}\Bigr)\sin\Bigl(\frac{\Delta_I}{2}\Bigr)
\end{equation*}
with parameters $\alpha_1 = 1.13+1.13\iu$, $\beta_1 = 1.13-1.13\iu$, 
$\Delta_R = \log\big(\frac{|\alpha_1|^2 |\beta_1|^2}{9}\big)$
and 
$\Delta_I = -\frac{\iu}{2} \log\big(\frac{\alpha_1 \beta_1}{\alpha_1^* \beta_1^*}\big)$.
This choice of double solitons is made to avoid a vanishing of the norm and energy quantities given by Eq.~\eqref{energy01}.

In our experiments, we chose the following system parameters: $L=6$, the discretization parameters $\Delta x=0.014$ and $\Delta t=0.002$.
The evolution of the left and right traveling solitons is shown in Fig.~\ref{fig:contour}, 
from which it can be seen that the solitons leave the computational domain almost without any reflection.

Following our simulation, we plot the time dependence of the energy of the solitons restricted to the computational domain. 
For this purpose, we discretize the energy in Eq.~\eqref{energy01} by
\begin{multline}
   E_n = \frac{1}{2\Delta x} 
   \sum_{j=1}^{J-1}\bigg[  
   (q_{j+1}^n-q_{j-1}^n)(\overline{q_{J-j-1}^{n}}-\overline{q_{J-j+1}^{n}}) \\+ 
\Delta x^2 (q_j^n \overline{q_{J-j}^{n}})^2
\bigg].
\end{multline}

\begin{figure}[t!]    
\includegraphics[width=90mm]{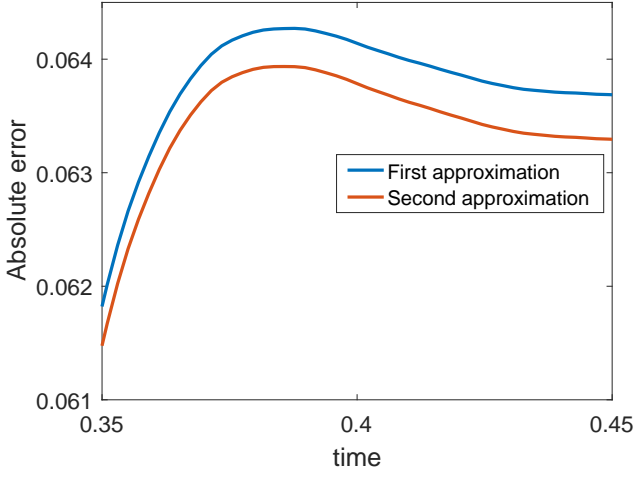}
\caption{The plot of the absolute error $||\mathcal{E}||_2^2$ versus time. 
The considered time interval is the period of interest, when solitons encounter boundaries.}\label{fig:error}
\end{figure}

The time dependence of the energy in the computational domain is shown in Fig.~\ref{fig:energy}. 
This plot shows that the energy vanishes, which means that there are almost no (reflected) waves as time passes.  

Finally, we show that the inclusion of the second-order approximation term in the TBC can slightly improve the results. 
To this end, we calculate the absolute error $\mathcal{E}$, i.e., the difference between the numerical solution with TBC imposed at $x=\pm L$ and the numerical solution for the extended interval $[-2L,2L]$ (such that boundaries are not reached within the considered time frame) restricted to $[-L,L]$, measured with the $L^2$-norm (discretized by the trapezoidal rule)
\begin{multline}
    ||\mathcal{E}||_2^2 = \Delta x \Bigg[ \sum_{j=1}^{J-1} 
    \Delta q_j^n \Delta \overline{q_{J-j}^{n}} + \\
    \frac{1}{2}(\Delta q_0^n \Delta \overline{q_{J}^{n}}+\Delta q_J^n \Delta \overline{q_{0}^{n}}) \Bigg],
\end{multline}
where $\Delta q_j^n = q_{j}^n - Q_{j}^n$ and $Q_{j}^n$ is the numerical solution for the extended interval. 
This is done to exclude the discretization error caused by the finite-difference scheme and to compare only errors caused by the approximations of TBCs.
The plot of this error versus time for the time period of solitons' leaving is shown in Fig.~\ref{fig:error}.

\begin{figure}[t!]
\includegraphics[width=90mm]{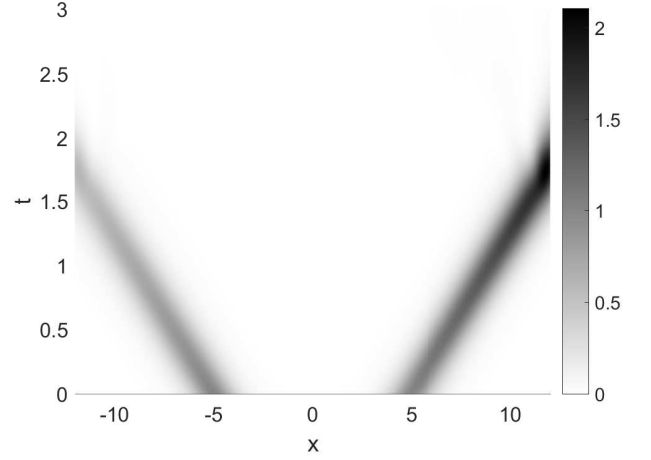}
\caption{Evolution of the asymmetric soliton profile given by Eq.~\eqref{in_con_2} with TBCs imposed at both (left and right) end points.}
\label{fig:contour2}
\end{figure}

\begin{figure}[th!]
\includegraphics[width=90mm]{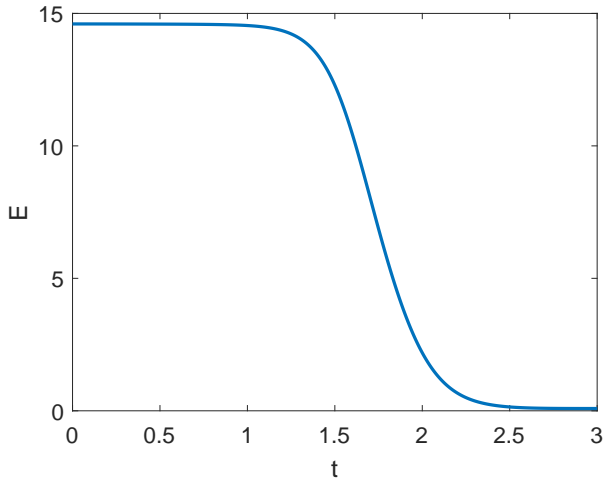}
\caption{Energy vs.\ time in the finite domain $x\in [-L,L]$ for the case of an asymmetric soliton evolution.}\label{fig:energy2}
\end{figure}

Here we also consider the asymmetric case, i.e.\ the case of asymmetric solitons. 
To do this, we use variational solutions in the numerical solution of the NNLS equation with TBC as initial conditions, which are described in Ref.~\cite{Hadi2019}:
\begin{gather}
    q(x,0)=\sum_{j=1}^2 q_j(x,0), \nonumber\\
    q_j(x,0) = A_j \exp(\iu B_j) \,\sech[C_j(x-X_j)] \nonumber\\
    \times\exp\bigl[\iu D_j(x-X_j)^2+ \iu E_j(x-X_j)\bigr],\label{in_con_2}
\end{gather}
where $A_1=A_2=1$, $B_1=0$, $B_2=0.1$, $C_1=C_2=1$, $D_1=D_2=0$,  $E_1=-E_2=2$, $X_1=-X_2=5$.
The evolution of the  traveling asymmetric soliton is shown in Fig.~\ref{fig:contour2}.
It shows similar dynamics as the symmetric counterpart, i.e.\ no reflection is visible. 
As an additional confirmation of the reflectionless propagation of the asymmetric soliton, we have plotted in Fig.~\ref{fig:energy2} the time dependence of the energy confined to the finite interval $[-L,L]$. The plot shows that TBC also works in these cases.

%%%%%%%%%%%%%%%%%%%%%%%%%%%%%%%%%%%
\section{Conclusions}
In this work, we have derived transparent boundary conditions (TBCs) for the nonlocal nonlinear Schr\"odinger equation using the so-called potential approach. 
Such boundary conditions allow to obtain the solution of an initial value problem given in an interval, which is equal to the solution of the problem for the whole space confined in this interval. 
The discretization of the derived TBCs and their numerical implementation were presented in detail. 

The confirmation of the nearly reflectionless transition by artificial boundaries was achieved by the simulation of traveling solitons. 
The time dependence of the energy in the computational domain was calculated to verify the obtained results. 
Although the first-order approximation of the TBC shows good results, the additional second-order terms are also considered to show that the absolute error decreases in this case (which is natural).
Although TBCs are commonly used for numerical simulations, they can also be explained from a physical point of view: 
The incoming wave does not ``feel'' the boundary where the TBC is imposed, which ensures that there is no or minimal loss in the transmission of waves from one domain to another. 

The above model can be used for the development and design of PT-symmetric optical waveguides that allow quasi-reflectionless propagation of solitons. 
The practical application of such functional materials in optoelectronic devices would allow to save resources and improve performance by reducing signal losses. 
Finally, we note that the above consideration, similar to \cite{Jambul1}, can be directly extended to optical waveguide networks by determining physically relevant conditions for the transparency of the branching points of the network. 
 Such structures are even more attractive from the point of view of optoelectronic applications. 
 A corresponding study is currently in progress.

%%%%%%%%%%%%%%%%%%%%%%%
\begin{acknowledgments}
The work is supported by the grant of the Ministry for Innovation Development of Uzbekistan (Ref. No. F-2021-440). 
One of the authors (DM) thanks the Associates Program of the Abdus Salam ICTP for his hospitality during his visit.
\end{acknowledgments}

%%%%%%%%%%%%%%%%%%%%%%%%%%%%%%%%%%%% Refs

\end{document}